\documentclass[12pt,twoside,pointlessnumbers,smallheadings]{revtex4}
\usepackage{amsmath}
\usepackage{amssymb}
\usepackage{color}
\usepackage{natbib}
\usepackage{hyperref}

\usepackage{hangcaption,fancybox,feynmp}
\usepackage{indentfirst}
\usepackage{graphicx}
\usepackage{epsfig,subfigure,psfrag}
\usepackage{CJK}
\usepackage{mathrsfs}
\usepackage{slashed}
\usepackage{url} % hyperref works too
\begin{document}

%%%%%%%%%%%%%%%%%%%%%%%%%%%%%%%%%%%

\title{On Extra Neutral and Charged Higgs bosons at the LHC
}
\author{Xiao-ping Wang $^{1}$,  Shou-hua
Zhu$^{1,2,3}$}

\affiliation{ $ ^1$ Institute of Theoretical Physics $\&$ State Key
Laboratory of Nuclear Physics and Technology, Peking University,
Beijing 100871, China \\
$ ^2$ Center for High Energy Physics, Peking University, Beijing 100871, China \\
$ ^3$ Collaborative Innovation Center of Quantum Matter, Beijing, China }

\date{\today}

\maketitle

\begin{center}
{\bf Abstract}
\end{center}

The properties of 125 GeV new particle, which was discovered in 2012 at the Large Hadron Collider (LHC), are found to be consistent with  those of the Higgs boson in the standard model (SM). Hereafter the new particle is dubbed as SM-like Higgs boson. However there is still spacious
room for physics beyond the SM (BSM) due to the limited energy and luminosity of the LHC. With more data, experiments will scrutinize whether the new particle is indeed the SM one or not. At the same time, one believes
that discovery of the SM-like Higgs boson is just the start of the new era of particle physics. The predominant topic is whether there are others new Higgs bosons as speculated in various BSM models. In this short review we will
 describe the current status of Higgs physics at the LHC and several BSM models which contain more Higgs sectors. In literature there are numerous studies on extended Higgs sector and a comprehensive review is beyond the scope
 of this review. Instead, we will present two latest studies on Higgs physics: (1) how to detect the charged Higgs boson and measure $\tan\beta$ after including the top polarization information, and (2) how to
 discover the extra neutral Higgs boson via the pair production of SM-like Higgs boson.

%\renewcommand{\baselinestretch}{1.2}
%\fontsize{12pt}{12pt}\selectfont


\newpage

\section{Introduction}

The discovery of standard model (SM) like Higgs boson~\cite{Aad:2012tfa,Chatrchyan:2012ufa} at the LHC is a milestone for understanding electroweak symmetry breaking.
%Besides the mass measurements of the SM-like Higgs boson, the measurements of its couplings to other SM particles are necessary to determine whether the SM-like Higgs boson is indeed the SM one or not.
The SM can successfully describe the strong and electroweak interactions, and it also provides the mechanism responsible for mass generation, i.e. by introducing one Higgs doublet. Although it is a very successful model, there are many reasons for
physics beyond the SM (BSM). Sometime BSM is also named as new physics (NP).
Theoretically the Higgs boson suffers from the notorious hierarchy problem \cite{Zhu:2012yv,Hu:2013cda}. The Higgs boson mass is a free parameter in the SM and could be enormous.
There are many solutions to this hierarchy problem, such as supersymmetry (SUSY) and little Higgs models etc. %The popular solution to this can be SUSY, Little Higgs model and top partner models.
 There are other motivations
 for BSM like neutrino mass, dark matter etc. In this paper we pay our attention on the Higgs related models. Typically the BSM will introduce more Higgs sectors. Searching such extended Higgs sector is an important
 task for collider experiments.

In order to seek the BSM, roughly speaking, there are two ways. The first way is to measure the Higgs properties as precise as possible. Any deviation from the SM predictions will be the indication of the BSM. The second way is to search for new particles not in the SM, especially the extra charged and/or neutral Higgs bosons. It is believed that there are likely rich Higgs sectors at weak scale.% \cite{Lee:1977eg,Gunion:1995zu}.
In this paper, after briefly describing the status of Higgs physics at the LHC and the BSM theory, we will present two recent studies on extra Higgs bosons: (1) detecting charged Higgs boson and measuring the important parameter $\tan\beta$ after including the top quark polarization information, and (2) discovering an extra neutral Higgs boson via pair production of SM-like Higgs bosons.

\section{Status on Higgs Physics at the LHC}

\begin{figure}[bht]
\begin{center}
\includegraphics[width=0.4\textwidth]{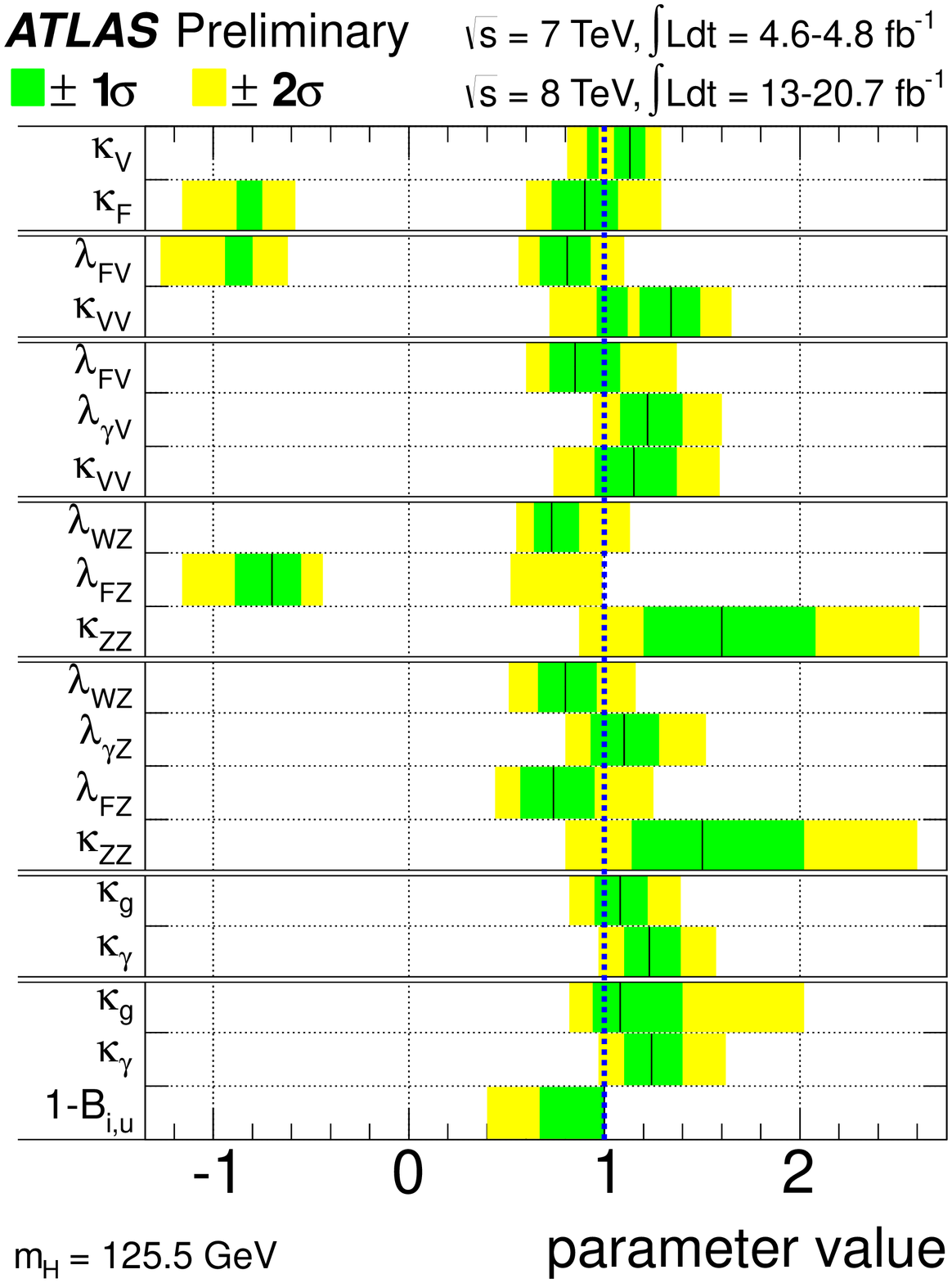}
\includegraphics[width=0.53\textwidth]{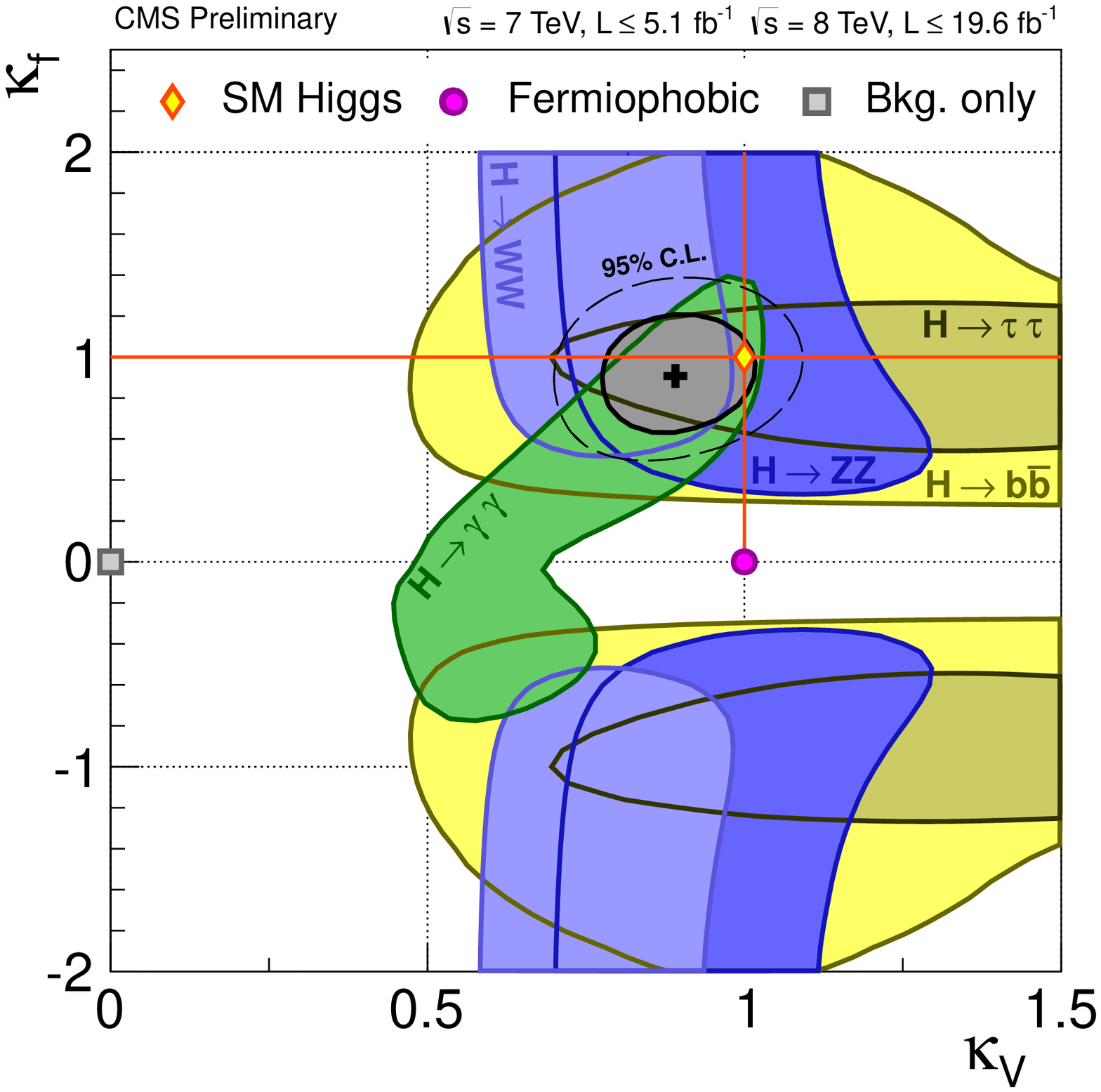}
\end{center}
\caption{\it The left figure is the summary of coupling scale factors for $m_H$=125.5 {\rm GeV} by ATLAS Collaboration~\cite{ATLAS:2013sla}. The best-fit values are represented by the solid black vertical line. The different parameter set in the different benchmark models, separated by double lines in the figure, are strongly correlated. In a sense they stand for the different parameterizations for the same data. The right figure is 68\% CL contours for the test statistic in the ($\kappa_V$ versus $\kappa_F$) plane for the individual channels (colored regions ) and the overall combination (solid thick lines) by CMS Collaboration~\cite{CMS:yva}.The thin dashed lines show the 95\% CL range for the overall combination. The black cross indicates the global best-fit values. The diamond show the SM Higgs boson point ($\kappa_V$, $\kappa_F$)=(1,1).}  \label{fig:ATLASCMS}
\end{figure}

There are many  measurements at the LHC with $\sqrt{s}=7$ TeV and integrated luminosity $\mathcal{L}=4.8$ $\rm{fb^{-1}}$, and with $\sqrt{s}=8$ TeV and $\mathcal{L}=20.7$ $\rm {fb^{-1}}$ for the most sensitive channels $H\rightarrow \gamma\gamma$, $H\rightarrow ZZ^*\rightarrow 4l$ and $H\rightarrow WW^*\rightarrow l\nu l\nu$  by ATLAS Collaboration~\cite{ATLAS:2012wma,ATLAS:2013sla}.  The CMS Collaboration also analyzed the same channels at $\sqrt{s}=7$ $\rm{TeV}$ with $\mathcal{L}=5.1$ $\rm{fb^{-1}}$ and $\sqrt{s}=8$ $\rm{TeV}$ with $\mathcal{L}=20$ $\rm{fb^{-1}}$~\cite{CMS:yva}.  Their results are summarized in Fig.~\ref{fig:ATLASCMS}, where $\kappa_i$ is the coupling scale factor. The $\kappa_i^2$ is defined as that ratio between the cross section $\sigma_{ii}$ or the partial decay width $\Gamma_{ii}$ ($i$ represents the specific SM final state) and the corresponding SM prediction. Taking the process $gg\rightarrow H\rightarrow \gamma\gamma$ as an example, we can write
\begin{equation}
\sigma\cdot BR(gg\rightarrow H\rightarrow \gamma\gamma)=\sigma_{SM}(gg\rightarrow H)\cdot BR_{SM}(H\rightarrow \gamma\gamma)\cdot\frac{\kappa_g^2\cdot\kappa_\gamma^2}{\kappa_H^2},
\end{equation}
where $\kappa_i=1$ means the measured coupling equals to the SM value. The scale factor $\kappa_F$ is the fermion factor with $\kappa_F=\kappa_t=\kappa_b=\kappa_\tau$, and $\kappa_V$ is the vector scale factor with  $\kappa_V=\kappa_W=\kappa_Z$. $\lambda_{ij}$ is the ratio of $\kappa_i$ and $\kappa_j$.
From the figure, we can see that no significant deviations have been observed! Note that the current data is only sensitive to the couplings between Higgs boson and gauge bosons/heavy fermions. In order to fully test SM, especially the Higgs
 potential, the measurements of Higgs self-coupling are needed. There are many theoretical works on this topic~\cite{Djouadi:1999gv,Djouadi:1999rca,Baur:2002rb,Baur:2002qd,Baur:2003gp,
Dolan:2012rv,Baglio:2012np,Dolan:2013rja}, and LHC can probe such coupling only after accumulating more data.

In the SM, the accumulated Higgs data came mainly from gluon-gluon fusion processes. Besides the gluon fusion channel, other Higgs boson production mechanisms also need to be tested. The Higgs associated production with vector boson (VH) has not been seen in the experiments~\cite{TheATLAScollaboration:2013hia,TheATLAScollaboration:2013lia}. The vector boson fusion (VBF) production channel is also very interesting while challenging to study. The observed significance of VBF channel is quite low which is only $1.3 \sigma$~\cite{TheATLAScollaboration:2013mia}. The Higgs associated production with top is important to measure top-Higgs coupling directly. This production channel has been studied for Higgs decay to diphoton, multi-leptons and bottom quark, and no signal has been observed yet ~\cite{TheATLAScollaboration:2013mia, Hig13020TWiki, Hig13019TWiki}.

In order to pin down whether the new particle is the SM Higgs boson or not, the measurement of Higgs spin-parity is necessary. ATLAS and CMS have already done such measurement based on the data of LHC with $ \sqrt s = 7,
8 \rm{TeV}$, $\mathcal{L}\sim 20\rm{fb^{-1}}$~\cite{Chatrchyan:2012jja,Aad:2013xqa}. The data are compatible with the SM prediction, namely $J^P = 0^+$. The ATLAS and CMS excluded $J^P = 0^-; 1^+; 1^-; 2^+ $ at confidence levels above $97.8\%$.

Besides the above measurements, LHC has also searched for other Higgs bosons. ATLAS has searched for another SM-like Higgs boson with the different mass~\cite{TheATLAScollaboration:2013zha} at $\sqrt{s}=8$ $\rm{TeV}$ with integrated luminosity $\mathcal{L}=20.7$ $\rm{fb^{-1}}$. They excluded a SM-like  Higgs boson in the mass range $260$ $\rm{GeV}\leq m_H\leq \rm{642~GeV}$ at 95\% confidence. Other searches for the neutral and charged Higgs bosons in supersymmetric (SUSY) model and two-Higgs doublet model (2HDM)~\cite{ATLAS:2013pma, Aad:2012cfr, TheATLAScollaboration:2013wia} have been carried out, and no positive signal has been found. They gave the constraints on the $m_A$ and $\tan{\beta}$ for those models. The CMS Collaborations did the similar analysis~\cite{CMS:2013hja,Chatrchyan:2012ya}, and they also found no deviations from SM backgrounds.

One more interesting direction is searching for the exotic decay of the SM-like Higgs boson. For various BSM models, Higgs boson could decay into particles not in the SM zoo. In Ref. \cite{ATLAS:2013pma}, the ATLAS collaboration searched for Higgs decay into invisible particles (e.g. dark matter) in the ZH associate production with Z decay leptonically. No deviation from SM expectation has been observed. CMS collaboration has also done similar analysis ~\cite{CMS:1900fga, CMS:2013yda} for ZH production channel, with Z decay hadroniclly and leptonically. A combination with VBF production has already been searched by CMS in \cite{CMS:2013bfa}, and no positive signal has been seen.

\section{Extra Higgs Boson in BSM Models}

Besides one doublet in the SM, BSM usually contains more scalar fields.
In 2-Higgs doublet model (2HDM) \cite{Gunion:1989we,Gunion:2002zf}, one more doublet is added into the model. Two Higgs doublets are also required in minimal SUSY models, because the superpotential must be a holomorphic function of the chiral supermultiplets and one is not allowed to take a complex conjugation of them as in the SM.
Moreover this can also keep the theory
anomaly free.  After symmetry breaking for the CP conserving case, there are five physical Higgs: two CP-even Higgs bosons ($h^0$, $H^0$), one CP-odd Higgs boson ($A^0$) and two charged Higgs bosons ($H^{\pm}$). In general, there are just six parameters in  Higgs sector: %~\cite{Chang:2013ona}
$m_{h^0}$, $m_{H^0}$, $m_{A^0}$, $m_{H^{\pm}}$, $\alpha$ and $\tan{\beta}$,  where $\alpha$ is the mixing angle between two CP-even Higgs ($h^0$ and $H^0$) and $\tan{\beta}$ is the ratio of vacuum value of  the two Higgs doublets ($\tan{\beta}=v_1/v_2$) related to the Yukawa couplings. There are many types of 2HDM depending on how the fermions couple with the Higgs sectors. The most popular types are Type-I and type-II. In Type-I 2HDM, one Higgs doublet couples to both up and down-type fermions and the other Higgs does not couple to fermions. In Type-II 2HDM, one Higgs doublet couples to up-type fermions and the other Higgs doublet couples to down-type fermions.
There are numerous studies in 2HDM.
Recently we found that CP spontaneous breaking seems intimately connected to the lightness of Higgs boson \cite{Zhu:2012yv,Hu:2013cda}. The phenomenology of this model is similar to the popular 2HDM. How to distinguish different 2HDM is an interesting topic.
%We are looking forward to see if there are extra Higgs in the LHC search.

SUSY is a symmetry between bosons and fermions~\cite{Barger:1993vc,Barger:1993gt,Zwirner:1993hq,Dine:1992yw,Wess:1992cp,
Djouadi:2008uw}. It cancels the quadratic loop divergence of Higgs mass by adding the SM particle partners, which solves the hierarchy problem in an elegant way. In the minimal supersymmetric standard model (MSSM) which is a
highly constrained Type-II 2HDM, there are two Higgs doublets and predict five physical Higgs bosons . By the remarkable idea of relating bosons and fermions, it also requires that there are a gluino ($\tilde{g}$), the super partner of gluon, squark ($\tilde{q}$), partners of all SM quarks, two partners of $W^{\pm}$ and $H^{\pm}$ named chargino ($\tilde{\chi}^2_{1,2}$), and partners of $Z$, $\gamma$ and Higgs named ($\tilde{\chi}^0_{1,2,3,4}$) where $\tilde{\chi}^0_1$ is the lightest supersymmetric particle (LSP).
The Higgs sector is very similar to Type-II 2HDM, but with less parameters. The tree level mass of Higgs in the MSSM is less than the Z mass. The 125 GeV Higgs mass provokes the naturalness problem in the MSSM. To have a 125 GeV Higgs mass, one solution is taking stop heavier to get greater loop contribution \cite{Arbey:2011ab}. But one can not set stop too high, or it will create another hierarchy between the Higgs mass and stop mass. There are solutions to this issue, such as the next minimal supersymmetric standard model (NMSSM)~\cite{Maniatis:2009re} and ``sister-SUSY"~\cite{Alves:2012fx}, which adding new scalars to modify the tree level mass relation of Higgs. Compared to the discovered Higgs boson with 125 GeV,
the mass of lightest supersymmetric Higgs boson is too light. Contrary to SUSY case, the dynamical electro-weak symmetry broken models usually lead to heavier Higgs boson.

There is another possibility to solve this problem by introducing the well designed global symmetry, as in the little higgs (LH) model~\cite{Schmaltz:2005ky,Schmaltz:2004de,Reuter:2013iya}. In LH model, the light Higgs boson is treated as the  pseudo-Nambu-Goldstone boson of the global symmetry.  As a realization, the littlest Higgs model is a minimal model with a global symmetry $SU(5)$. After symmetry breaking, there are 14 Goldstone bosons: Four of them are eaten leading to four massive vector bosons (a $SU(2)$ triplet $Z_H$, $W^{\pm}_H$, and a $U(1)$ boson $Z_H$). Ten of them are remained as Goldstone bosons transformed under the SM gauge group as a doublet $h$ (which is the SM Higgs doublet) and a triplet $\phi$. In order to cancel the quadratic divergence of Higgs mass from top-quark, it needs a vector-like pair of colored Weyl fermions which leading to a new heavy vector-like quark with charge +2/3~\cite{Dobado:2009sc}.

To summarize, there are numerous BSM models which usually introduce the extended Higgs sectors. Discovering/excluding the extra charged and/or neutral Higgs bosons at high energy collider is crucial to confirm/exclude the corresponding BSM models.

\section{Searching for charged Higgs boson in polarized top-quark}

In the SM, there is only one neutral Higgs boson. Therefore the charged Higgs boson is unambiguous signature of the BSM. In fact charged Higgs boson is quite common in many new physics as discussed in previous section. In our recent work~\cite{Cao:2013ud}, we concentrated on the type-II 2HDM, where one Higgs doublet couples to up-type fermions and the other Higgs doublet couples to down-type fermions.

In type-II 2HDM, the coupling among charged Higgs boson and quarks can be written as
\begin{equation}
g_{H^-\bar{d}u}=\frac{g}{\sqrt{2}m_W}(m_d\tan{\beta}P_L+m_u\cot{\beta}P_R),
\label{feynman_tbh}
\end{equation}
where $P_{L/R}=(1\mp\gamma_5)/2$ is the chirality projector.
Here $\tan{\beta}$ is a crucial parameter of 2HDM which is defined as the ratio between the vacuum expectation values of two Higgs doublet.
As indicated in Eq.\ref{feynman_tbh}, the coupling strength is proportional to the fermion's mass. We can consider the most massive fermions: the third generation. Thanks to its heavy mass, the top-quark decays promptly so that the chirality information of top can be kept in its decay products. With this idea, we can use the decay products to reconstruct the top-quark and its polarization information, and measure $\tan{\beta}$ with better precision.

 Charged Higgs in type-II 2HDM can be produced in three ways: (1) $pp\rightarrow \gamma/ Z\rightarrow H^+H^-$; (2) $gb\rightarrow tH^-(\bar{b}g\rightarrow \bar{t}H^+) $; and (3) $q\bar{q}^{\prime}\rightarrow W\rightarrow AH^{\pm}/hH^{\pm}/HH^{\pm}$. The cross section of process (1) decreases with $m_{H^{\pm}}$ more rapidly compared to the other two, and the process (3) contains other unknown parameter $m_A$. In our work, we limited ourself on the $tH^-/\bar{t}H^+$ associated production~\cite{Baglio:2011ap,Godbole:2011vw,Gong:2012ru,Beccaria:2013yya,Gong:2014tqa}
\begin{equation}
gb\rightarrow tH^-\rightarrow t\bar{t}b
\label{signal_1}
\end{equation}
\begin{equation}
g\bar{b}\rightarrow \bar{t}H^+\rightarrow \bar{t}t\bar{b}.
\label{signal_2}
\end{equation}

These two process both generate one top-quark, one anti-top-quark and a bottom-quark which can not be distinguished by the final states. If we want to probe the $H^+\bar{t}b$ coupling, we should take care of the different origin of the anti-top in two process: the $\bar{t}$ of process (\ref{signal_1}) is the decay product of charged Higgs, while the $\bar{t}$ in process (\ref{signal_2}) is associated produced with the charged Higgs. We calculate the helicity amplitudes of different processes and obtain their degree of top polarization respectively as \cite{Berger:2012an,Berger:2011hn},
\begin{eqnarray}
D_{\rm decay}&\equiv&\frac{\Gamma(\bar{t}_L) - \Gamma(\bar{t}_R)}{\Gamma(\bar{t}_L) + \Gamma(\bar{t}_R)} = \frac{(m_t \cot\beta)^2 - (m_b\tan\beta)^2}{(m_t\cot\beta)^2 + (m_b\tan\beta)^2}\\
D_{\rm prod}(\hat{s}) &\equiv& \frac{\hat{\sigma}(t_R)-\hat{\sigma}(t_L)}{\hat{\sigma}(t_R)+\hat{\sigma}(t_L)} = \frac{(m_t\cot\beta)^2-(m_b \tan\beta)^2}{(m_t\cot\beta)^2+ (m_b \tan\beta)^2 } \times \hat{R}_{\rm prod}.
\label{dilution_def}
\end{eqnarray}

\begin{figure}[bht]
\begin{center}
\includegraphics[width=0.4\textwidth]{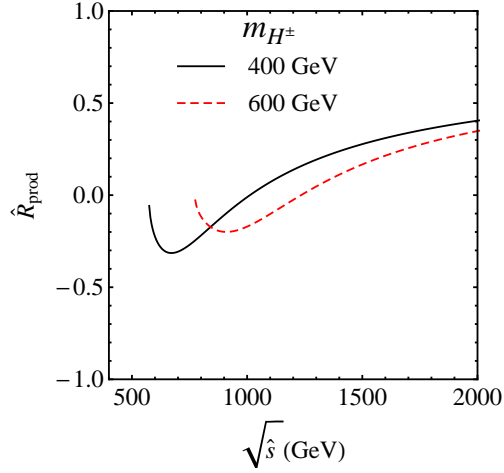}
\end{center}
\caption{\it The dilution factor in Eq.~\eqref{dilution_def} as a function of the energy of the overall c.m. frame ($\sqrt{\hat{s}}$) with $m_{H^\pm}=400\rm {~ GeV}$(solid black) and $m_{H^\pm}=600\rm{~GeV}$ (dashed red){\color{red}\cite{Cao:2013ud}}. }\label{fig:Rprod}
\end{figure}

The distribution of dilution factor $\hat{R}_{\rm prod}$ as a function of the c.m. energy $\sqrt{\hat{s}}$ of the hard scattering process for $m_{H^{\pm}}=400$ GeV and   $m_{H^{\pm}}=600$ GeV is shown in Fig.~\ref{fig:Rprod}. From the figure, we can find the absolute value of the dilution factor is less than 0.5 and the sign of the dilution factor can turn from negative in the threshold region of the $tH^-$ to positive in the large invariant mass region. The sign of the dilution factor is very important for the measurement of top polarization. Therefore, we focus our attention on the anti-top from the $H^-$ decay rather than the associated production with the $H^+$.

Since we searching for the events with one top-quark, one anti-top-quark and a bottom/anti-bottom-quark which can not be distinguished by the final states both, $gb\rightarrow tH^-$ and $g\bar{b}\rightarrow \bar{t}H^+$ process can contribute to the signal. Since we are interested in the polarization of anti-top from the charged Higgs decay, we call $g\bar{b}\rightarrow \bar{t}H^+$ as irreducible background. Here we have two tasks: one is to distinguish the $gb\rightarrow tH^-$ from $g\bar{b}\rightarrow \bar{t}H^+$, and the other is to suppress the SM backgrounds. We demand $\bar{t}\rightarrow l^-\bar{b}\bar{\nu}$ and $t\rightarrow bjj$. The anti-top polarization can be
inferred via the angular distribution of charged leptons as described below. The signal and backgrounds are :
\begin{eqnarray}
\begin{split}
Signal:&gb \rightarrow tH^-\rightarrow (W^+b)(\bar{t}b) \rightarrow (jjb)(l^-\bar{\nu}\bar{b}b)\\
Irreducible~background:&g\bar{b} \rightarrow \bar{t}H^+\rightarrow (W^-\bar{b})(t\bar{b}) \rightarrow (l^-\bar{\nu}\bar{b})(jjb\bar{b})\\
SM Background:&pp\rightarrow t\bar{t}j_b\rightarrow bW^+\bar{b}W^-j_b\rightarrow j_bj_bj_bjjl^-\bar{\nu}\\
&pp\rightarrow t\bar{t}j\rightarrow bW^+\bar{b}W^-j\rightarrow j_bj_bjjjl^-\bar{\nu},
\end{split}
\end{eqnarray}
where $j_b$ means b-jet contains b or $\bar b$. Note that our analysis in this work is at parton level. There are five jets in the final states. We order the five jets in the final states by their $p_T$ and examine their $p_T$ distributions~\cite{Cao:2013ud}. Based on these information we can introduce the basic $p_T$ cut conditions on signal and backgrounds as:
\begin{equation}
p_T(j_{\rm 1st}) \geq 120~{\rm GeV}, \qquad p_T(j_{\rm 2nd})\geq 80~{\rm GeV}, \qquad p_T(j_{\rm 3rd})>60~{\rm GeV}.
\label{hardcut}
\end{equation}
In order to suppress the huge SM background, we need to reconstruct the intermediate states. We use $\chi^2$-template method~\cite{Berger:2011xk} to reconstruct the $t\bar{t}$ pair and singles out the extra jet:
\begin{equation}
\chi^2 = \frac{(m_W - m_{jj})^2}{\Delta m_W^2} + \frac{(m_t - m_{j\ell^- \bar{\nu}})^2}{\Delta m_t^2} + \frac{(m_t - m_{jjj})^2}{\Delta m_t^2},
\end{equation}
where $\Delta m_x^2$ is mass width for x particle which is calculated in SM.  At the same time, we use the $W$-boson on-shell condition to reconstruct neutrino:
\begin{equation}
p_{\bar{\nu} L}=\frac{1}{2p_{\ell^- T}^2}\left[\left(m_W^2+2\overset{\rightarrow }{P}_{\ell^-T}\cdot\!\!\overset{\rightarrow}{\not{\!\rm E}}_T\right)p_{\ell^-L}\pm E_{\ell^-}\sqrt{\left(m_W^2+2\overset{\rightarrow }{P}_{\ell^{-}T}\cdot\!\!\overset{\rightarrow }{\not{\!\rm E}}_T\right)^2-4p_{\ell^-T}^2 \!\!\not{\!\rm E}_T^2}~\right]
\end{equation}
We also scrutinize the distribution of $m_{\bar{t}j_{extra}}$ and $p_T(j_{extra})$~\cite{Cao:2013ud} and introduce the cut conditions to distinguish $gb\rightarrow tH^-$ from $g\bar{b}\rightarrow \bar{t}H^+$:
\begin{equation}
\Delta M_{\bar{t}j_{\rm extra}}\equiv \left| M_{\bar{t} j_{\rm extra}}-M(H^\pm) \right| \leq 5{\rm ~GeV}, ~~~p_T(j_{\rm extra}) \geq 120~{\rm GeV}.
\end{equation}
Finally, we demand the extra jet to be a $b$ jet and choose the b-tag efficiency as 60\% and mis-tagging efficiency as 2\%.
The cut efficiency is listed in Tab. \ref{tbl-efficiency}.
We can get the significance of the signal well above $5\sigma$ for a broad range of $\tan{\beta}$. Even for $\tan{\beta}=6$, there are more than 300 events survive for $m_{H^\pm}=400$ GeV at the 14 TeV LHC with an integrated luminosity of 100 $fb^{-1}$.

\begin{table}
\begin{center}
\caption{Number of events of the signal and backgrounds at the 14 TeV LHC with an integrated luminosity of $100~{\rm fb}^{-1}$ for $m_{H^\pm}=400~{\rm GeV}$ and three values of $\tan\beta$. This table is taken from Ref. \cite{Cao:2013ud}. }
\label{tbl-efficiency}
\begin{tabular}{c||c|c|c|c|c|c||c|c}
\hline
$\tan\beta$ & \multicolumn{2}{c|}{1} & \multicolumn{2}{c|}{6} & \multicolumn{2}{c||}{40} & \multicolumn{2}{c}{SM backgrounds}\tabularnewline
\hline
 & $tH^{-}$ & $\bar{t}H^{+}$ & $tH^{-}$ & $\bar{t}H^{+}$ & $tH^{-}$ & $\bar{t}H^{+}$ & $t\bar{t}j$ & $t\bar{t}b$\tabularnewline
\hline
\hline
Inclusive rate & 23310 & 23300 & 1255 & 1227 & 24660 & 23520 & $1.075\times10^{7}$ & 234000\tabularnewline
Hard $p_T$ cuts & 11843 & 13466 & 687 & 719 & 14421 & 13890 & $2.12\times10^{6}$ & 25052\tabularnewline
$\Delta M_{\bar{t}j_{\rm extra}}$ & 4980 & 368 & 672 & 20 & 5680 & 383 & 39238 & 386\tabularnewline
$p_{T}(j_{{\rm extra}})$ & 3910 & 305 & 532 & 16 & 4375 & 310 & 14942 & 171\tabularnewline
$b$ tagging & 2346 & 183 & 312 & 10& 2625 & 186 & 299 & 102\tabularnewline
\hline
\hline
Number of events & \multicolumn{2}{c|}{2529} & \multicolumn{2}{c|}{322} & \multicolumn{2}{c||}{2811} & \multicolumn{2}{c}{401}\tabularnewline
$S/B$ & \multicolumn{2}{c|}{6.3} & \multicolumn{2}{c|}{0.8} & \multicolumn{2}{c||}{7.0} & \multicolumn{2}{c}{$-$}\tabularnewline
$S/\sqrt{B}$ & \multicolumn{2}{c|}{126.3} & \multicolumn{2}{c|}{16.1} & \multicolumn{2}{c||}{140.3} & \multicolumn{2}{c}{$-$}\tabularnewline
$\sqrt{S+B}$ & \multicolumn{2}{c|}{54.1} & \multicolumn{2}{c|}{26.9} & \multicolumn{2}{c||}{56.7} & \multicolumn{2}{c}{$-$}\tabularnewline
\hline
\end{tabular}
\end{center}
\end{table}

With the reconstructed $\bar{t}$, $j_{extra}$ and $H^-$ and enough significance, we can measure the anti-top's polarization. Here we define an angle between the charged lepton momentum in the rest frame of $\bar{t}$ to the anti-top momentum in the rest frame of $H^-$:
\begin{equation}
\frac{d\sigma}{\sigma d\cos\theta_{\rm hel}}=\frac{1}{2} (1+D\cos\theta_{\rm hel}).
\end{equation}
Given the distribution of $\frac{d\sigma}{\sigma d\cos\theta_{\rm hel}}$, we can get the polarization of the anti-top quark via
 \begin{equation}
D=3\sum_{i=1}^{10} \cos\theta_i \left(\frac{d\sigma}{\sigma d\cos\theta}\right)_i\Delta \cos\theta
 =\frac{3\sum_{i=1}^{10} \cos\theta_i N_i}{\sum_{i=1}^{10} N_i},
\label{d-definition}
\end{equation}
where $N_i$ means the rescaled event number of the ith bin in the distribution of $\frac{d\sigma}{\sigma d\cos\theta_{\rm hel}}$. In our analysis, there are only 10 bins distributed between $\cos\theta_{\rm{hel}}=-1$ and $\cos\theta_{\rm{hel}}=1$, so the i is from 1 to 10. For simplicity, we introduce the statistical error of the degree of anti-top polarization as following:
\begin{equation}
\Delta D=\sqrt{\sum_{i=1}^n\left|\frac{\partial D}{\partial N_i}\right|^2 \left({\Delta N_i}\right)^2}.
\end{equation}

\begin{figure}[bht]
\begin{center}
 \includegraphics[width=0.4\textwidth]{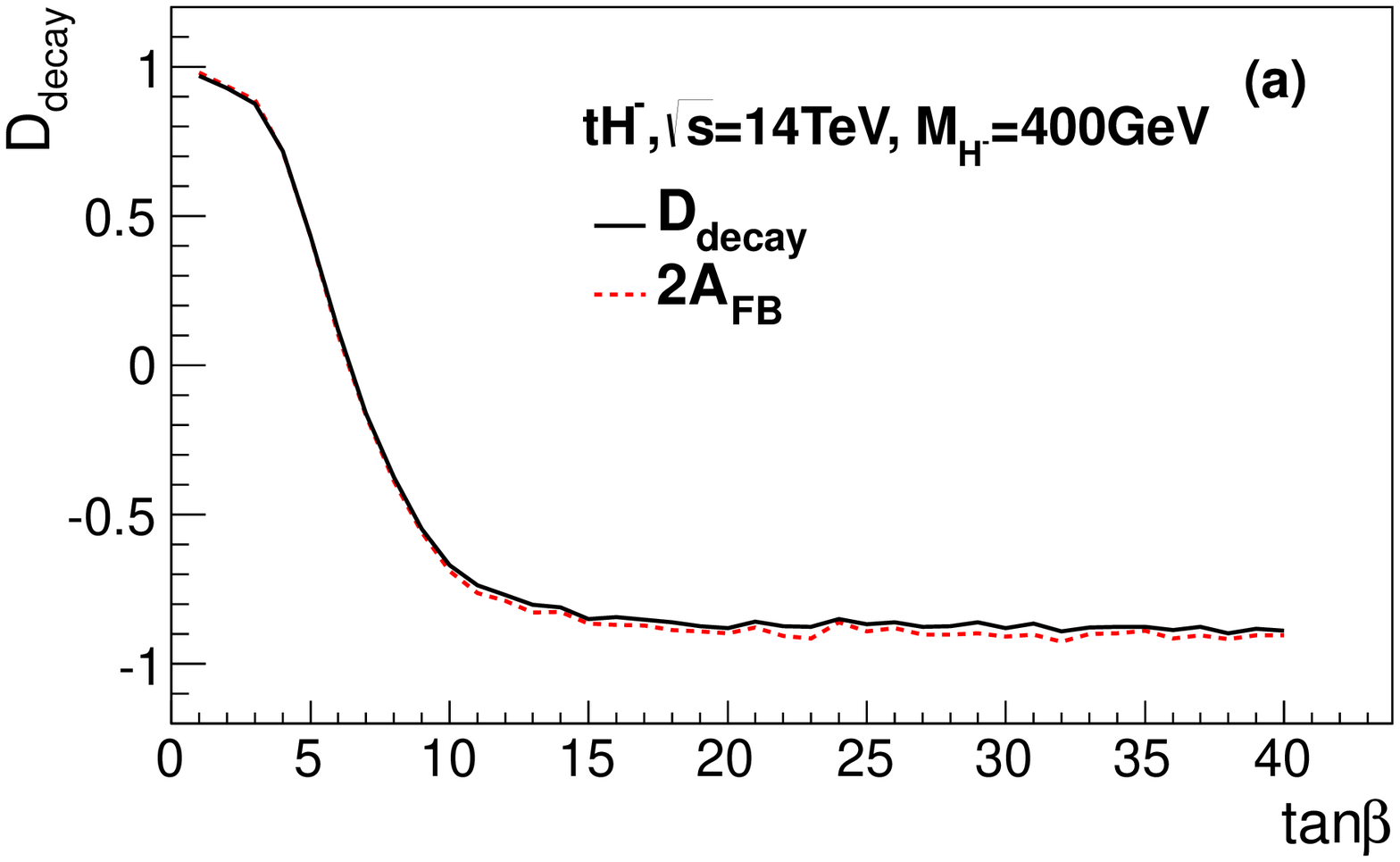}
  \includegraphics[width=0.4\textwidth]{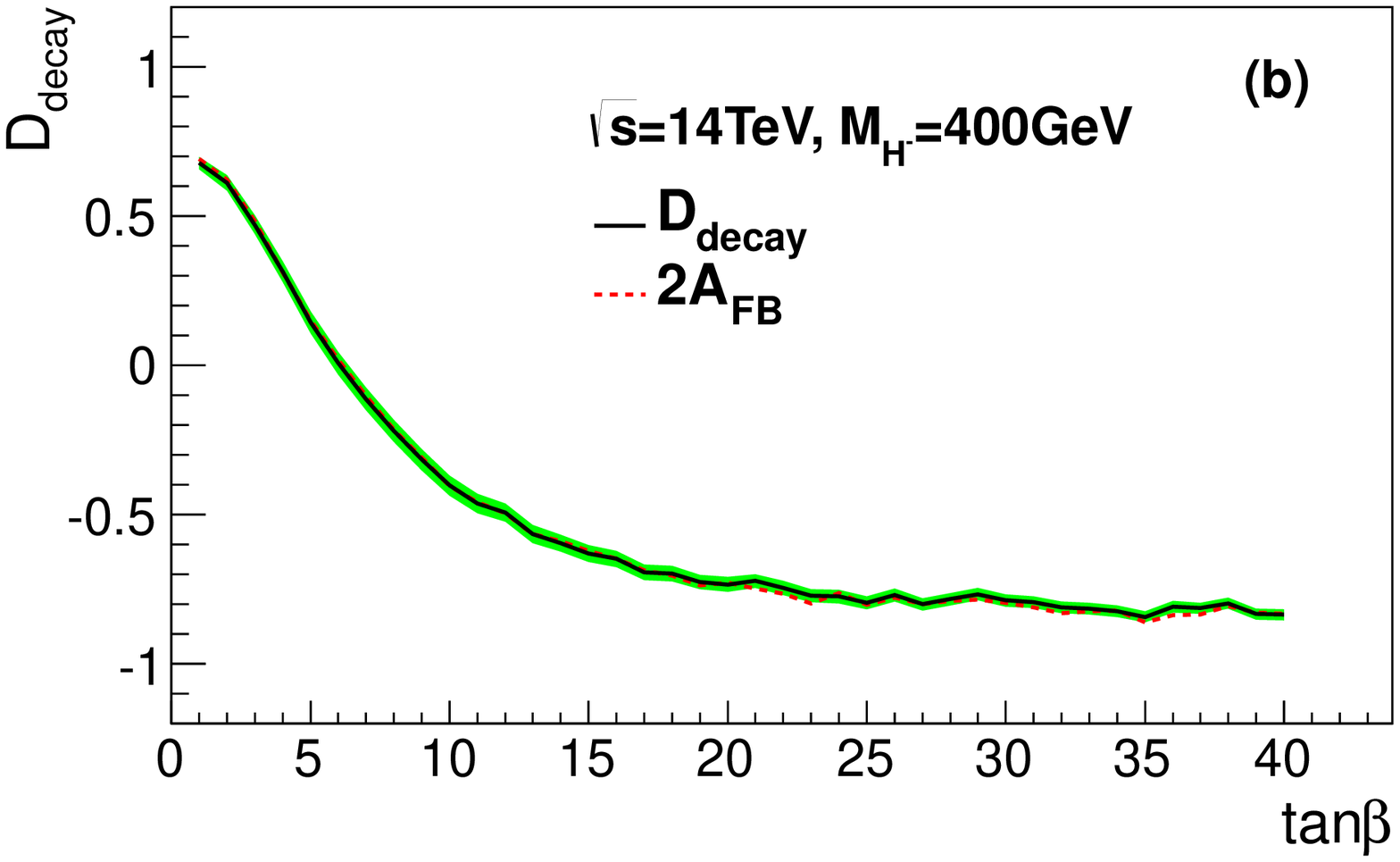}
  \end{center}
  \caption{\it (a) The degree of polarization of the anti-top quark as a function of $\tan\beta$ of the $tH^-$ signal event and (b) of all the signal and background processes with $m_{H^\pm}=400~\rm{ GeV}$. The solid black curve shows the degree of polarization defined in Eq.~\eqref{d-definition}; the dashed red curve shows $2A_{FB}$. The green band in (b) represents only the statistical uncertainties.}\label{fig:toppol}
\end{figure}

The polarization degree of the anti-top quark as a function of $\tan{\beta}$ is shown in Fig.~\ref{fig:toppol}. $A_{FB}\equiv\frac{\sigma_F -\sigma_B}{\sigma_F + \sigma_B}$ is also plotted for comparison. The anti-top quark polarization is a good probe for a wide range for $\tan{\beta}$, while the intermediate $\tan{\beta}$ is hard to measure. However, $D_{\rm decay}$ changes rapidly in the region of $\tan{\beta}=5\sim10$. This feature helps us to determine $\tan{\beta}$ with the top polarization information. Figure~\ref{fig:toppol}(b) tell us that the polarization can not reach $\pm1$ because of the SM backgrounds and the signal events loss by the cut conditions.

\section{ Detecting extra neutral Higgs boson via pair production SM-like Higgs bosons}

The discovery of the SM-like Higgs boson provides a new method for BSM search. In this section we will explore the capacity of discovering an extra neutral Higgs boson via pair production of SM-like Higgs bosons. In the SM, Higgs boson pair production is crucial to measure the Higgs triple coupling. Here the Higgs boson pair is mainly produced through triangle and box diagrams in gluon gluon fusion process. The cross section is about 30 fb on LHC 14 TeV~\cite{Goertz:2013eka,Ellwanger:2009dp, Goertz:2013kp, Shao:2013bz,Dawson:2006dm, Jin:2005gw, BarrientosBendezu:2001di, Dawson:1998py, Glover:1987nx}. A recent NNLO calculation for Higgs pair production suggest the cross section to be 40 fb~\cite{deFlorian:2013jea,deFlorian:2013uza}. Such small cross section implies limited statistics. Recently we consider the possibility of Higgs pair can be produced as the decay product of
 a new resonance~\cite{Liu:2013woa}. Compared to the case in the SM, the signal can be greatly enhanced ~\cite{Arhrib:2013oia,Baglio:2014nea}.

For simplicity, we analyze the process in the effective Lagrangian, which is written as
\begin{eqnarray}
\mathcal{L}=f_1\frac{\sqrt{2}\alpha_s}{12\pi v}SG^a_{\mu\nu}G^{a\mu\nu}+f_2\frac{(m_H)^2}{v}SHH.
\end{eqnarray}
Here scalar S can couple with colored particles in the loop and the coupling strength with gluon-gluon depends on specific models. The S can couple with SM-like Higgs H through a portal-like potential which provides the significant branching ratio for S decaying into HH. 
It should be noted that ATLAS and CMS looked for heavy Higgs bosons via their SM-like decay channels. In this work we assume that the other decay branching ratios of S are negligibly small.
To consider the constraints from experiments and give a allowed cross section for $pp\rightarrow S\rightarrow HH$, we discuss the resonant production of $S$ in the context of renormalizable coloron model \cite{Simmons:1996fz,Hill:2002ap,Bai:2010dj}. It should be noted that in 2HDM this resonant production process can occur for suitable parameters. In the coloron model, the strong gauge group is $SU{(3)_1} \times SU{(3)_2}$, which is broken down to $SU{(3)_S}$ in the standard model. The spontaneous breaking is provided by a complex scalar field $ \Phi $, which has representation of $ (3,\overline 3 ) $ and obtains a diagonal vacuum expectation value (vev). There is another complex scalar $\phi$ providing the spontaneous breaking of $SU(2)\times U(1)$, similar to that in the SM. Because of  mixing between $\Phi$ and $\phi$, there are two neutral Higgs particles S and H. Besides mass of SM-like Higgs: $m_H$ which is chosen as $125\rm{GeV}$, there are still seven parameters: two vev $\upsilon _h,\upsilon _\phi  $, one mixing angle $ \theta $, four masses $ \{ {m_S},{m_I},{m_{{G_H}}},{m_{G'}}\}  $ for heavy singlet scalar, singlet pseudo scalar, color octet scalar and coloron.
 Considering the constraints from experiments, we choose benchmark point as:
\begin{eqnarray}
\{ {\upsilon _h},{\upsilon _\phi },\sin \theta ,{m_S},{m_I},{m_{{G_H}}},{m_C}\}  = \{ 246\rm{GeV},4.2\rm{TeV},0.1,400\rm{GeV},1\rm{TeV},1\rm{TeV},3\rm{TeV}\}.
\end{eqnarray}

With this benchmark point, the cross section for signal process $ \sigma _{gg}^S \cdot BR(S \to hh) $ is $0.5$pb at LHC with $\sqrt s=14\rm{TeV}$. In our analysis, we use this number as the signal production cross section. The produced Higgs pair decay promptly. In order to examine the potential for different Higgs decay modes, we give the different BR of two Higgs final states according to Ref.~\cite{Dittmaier:2011ti} in Table.\ref{tab:BR}:
 \begin{table}[bht]
\centering
\begin{tabular}{lccccc}
\hline
decay mode &$b\bar{b}$ &WW &ZZ&$\gamma\gamma$&$\tau\tau$\\ \hline
$b\bar{b}$ &  $3.34 \times 10^{-1}$  & $1.25 \times 10^{-1}$&$1.54 \times 10^{-2}$& $1.33\times 10^{-3}$&$3.68\times 10^{-2}$\\
WW& $1.25 \times 10^{-1}$ & $4.67 \times 10^{-2}$ & $5.77 \times 10^{-3}$  & $4.97 \times 10^{-4}$&$1.38\times10^{-2}$\\
ZZ& $1.54 \times 10^{-2}$ & $5.77 \times 10^{-3}$ & $7.13 \times 10^{-4}$  & $6.14 \times 10^{-5}$&$1.7\times 10^{-3}$\\
$\gamma\gamma$&$1.33 \times 10^{-3}$& $4.97 \times 10^{-4}$ & $6.14 \times 10^{-5}$ &  $5.29 \times 10^{-6}$ &$1.46\times 10^{-4}$\\
$\tau\tau$&$3.68\times 10^{-2}$&$1.38\times10^{-2}$&$1.7\times 10^{-3}$&$1.46\times 10^{-4}$&$4.1\times10^{-3}$\\\hline
\end{tabular}
\caption{The product of two branching ratio of different decay modes for Higgs boson \cite{Dittmaier:2011ti}.}\label{tab:BR}
\end{table}

%Table.\ref{tab:BR} lists the measurement potential of different decay modes. If we want to consider the decay mode of $Z$ and $W^{\pm}$ bosons, we also should consider the BR of their decay products. In our analysis, we consider %the leptonic decay mode of gauge bosons which can suppress the huge QCD background compared to the hadronic decay modes.

From table.\ref{tab:BR} we can see that there are 15 combinations. The largest BR is the $b\bar{b}$ decay mode  but is also heavily polluted by QCD backgrounds.
In order to get reasonable signal cross section and controllable QCD backgrounds, we focus on one Higgs decay to $b\bar b$ while the other Higgs decay to $\tau^-\tau^+$. The lifetime of $\tau$ is so short that it promptly decays to leptons or hadrons. Here we analyze the hadronic decay of $\tau$. The signal and backgrounds contain:
\begin{eqnarray}
\begin{split}
\rm{sig}&:pp\rightarrow S\rightarrow HH\rightarrow\left(b\bar b\right)\left(\tau^+\tau^-\right)\\
\rm{bkg}&:pp\rightarrow b\bar b\tau^+\tau^-\\
&pp\rightarrow jj\left(Z\rightarrow \tau^-\tau^+\right)\\
&pp\rightarrow t\bar t\rightarrow \left(b\tau^+\nu\right)\left(\bar b\tau^-\nu\right).
\end{split}
\end{eqnarray}
\indent We generate the signal and background events in Madgraph with the default cut conditions. The events are put into pythia and PGS to do detector simulation. The corss section after the basic cut for signal is $0.038\rm{pb}$ and for backgrounds are: $\sigma(b\bar b\tau\tau)=2.57\rm{pb}$, $\sigma(jj\tau\tau)=125.88\rm{pb}$, $\sigma(t\bar t)=5.24\rm{pb}$. We require that each event contains at least two jets and two $\tau$ with $p_T\geq 20\rm{GeV}$. The cut effeciency for signal is 1/6 and for background is 1/10.
We then reorder the jets and hadronic $\tau$ with $p_T=\sqrt{p_x^2+p_y^2}$:
\begin{eqnarray}
\begin{array}{ll}
p_T(j_1)\geq p_T(j_2);&p_T(\tau_1)\geq p_T(\tau_2)
\end{array}
\end{eqnarray}
We define $H_T$ as $H_T=p_T(j_1)+p_T(j_2)+p_T(\tau_1)+p_T(\tau_2)$ and compare the different distribution of $p_T$ and $H_T$. In order to suppress backgrounds, we choose the basic cut conditions as
\begin{eqnarray}
\begin{array}{ccc}
p_T(j_1)\geq 100~\rm{GeV};&p_T(\tau_1)\geq 60~\rm{GeV};&H_T\geq 240~\rm{GeV}.
\end{array}
\end{eqnarray}

\indent In order to further suppress the huge backgrounds, we try to reconstruct the intermediate states and study their properties. There are two jets and two hadronic-decay $\tau$ which come from two H respectively. We use their information to reconstruct the momentum  of H and S. We can also obtain the invariant mass of $jj$, $\tau\tau$ and $jj\tau\tau$. Because H is boosted, the $p_T$, $\Delta R_{jj}=\sqrt{\Delta\eta_{jj}^2+\Delta\phi_{jj}^2}$ ($\eta$ is pseudo-rapidity and $\phi$ is azimuth angle), and $\Delta R_{\tau\tau}$ of signal are different from those of backgrounds. We choose the cut conditions as

\begin{eqnarray}
\begin{array}{lll}
p_T\left(H\right)\geq 100~\rm{GeV};&\Delta R_{j_1j_2}\leq 2.6;&\Delta R{\tau\tau}\leq 2.2\\
m_{ll}\in\left[100~\rm{GeV},150~\rm{GeV}\right];&m_{jj}\in\left[90~\rm{GeV},140~\rm{GeV}\right]
;&m_{Final}\geq 300~\rm{GeV}.
\end{array}
\end{eqnarray}
 \indent The cut condition reduces the signal by one order while reduces the backgrounds by 2-3 order. For b-jet, the b-tag efficiency is chosen as 0.6. For light flavor jet, it can be mis-tagged as the b-jet with the probability of  0.02~\cite{Aad:2009wy}. The effective cross sections for signal and backgrounds are $2.0\times10^{-4}~\rm{pb}$  and $3.96\times10^{-4}~\rm{pb}$ respectively. If the integrated luminosity is $1000~\rm{fb^{-1}}$, there are 200 signal events and 396 background events. The signal significance is $\frac{S}{\sqrt {S+B}}=8.2$ and the $\frac{S}{B}$ is 0.5. Our results show that di-Higgs production can be an excellent mode to discover such new resonance.

 \begin{table}[ht]
\small
\setlength{\tabcolsep}{0.1pt}
\centering
\begin{tabular}{l|c|c|c|c|c}
\hline
decay mode &$b\bar{b}$ &WW &ZZ&$\gamma\gamma$&$\tau\tau$\\ \hline\hline
$b\bar{b}$ &  $\begin{array}{c}N=2540,\frac{S}{B}=0.027\\
\frac{S}{\sqrt {S+B}}=8.3\end{array}$  & $\begin{array}{c}N=1570,\frac{S}{B}=0.058\\\frac{S}{\sqrt {S+B}}=9.5\end{array}$&$N\simeq0$& $\begin{array}{c}N=59,\frac{S}{B}=6.6\\\frac{S}{\sqrt {B+S}}=7.15\end{array}$&$\begin{array}{c}N=200,\frac{S}{B}=0.5\\\frac{S}{\sqrt{S+ B}}=8.61\end{array}$\\\hline
WW& - & $N\simeq0$ & $N\simeq0$  & $\begin{array}{c}N=1,\frac{S}{B}\sim10^{-3}\\\frac{S}{\sqrt {B}}=0.025\end{array}$&$\begin{array}{c}N=1,\frac{S}{B}=0.003\\\frac{S}{\sqrt {B}}=0.053\end{array}$\\\hline
ZZ& - & - & $N\simeq0$  & $N\simeq0$&$N\simeq0$\\\hline
$\gamma\gamma$&-& - & - &  $N\simeq0$ &$\begin{array}{c}N=3,\frac{S}{B}=0.5\\\frac{S}{\sqrt{ S+B}}=1\end{array}$\\\hline
$\tau\tau$&-&-&-&-&$\begin{array}{c}N=9,\frac{S}{B}=0.64\\\frac{S}{\sqrt {S+B}}=1.87\end{array}$\\\hline
\end{tabular}
\label{tab:discovery}
\caption{\small{The signal event number and significance of different signal at LHC with $\sqrt s=14\rm{TeV}$, $\mathcal{L}=1000\rm{fb^{-1}}$~\cite{Liu:2013woa}.}}
\end{table}

We have investigated 15 combinations for different SM-like Higgs decay modes and the final results are shown in table \ref{tab:discovery}. We did all analysis with the assumption $\sqrt s=14~\rm{TeV}$ and $\mathcal{L}=1000~\rm{fb^{-1}}$. There are two promising combination which can be excellent probes for an extra neutral Higgs S: $b\bar b\gamma\gamma$ and $b\bar b\tau^-\tau^+$. These processes not only have enough signal events but also have large signal significance and S/B ratio. For the other two less promising combinations: $b\bar bb\bar b$ and $b\bar b WW^*$, they have enough signal events and signal significance, but they don't have large enough S/B ratio due to the huge QCD backgrounds.

\section{conclusion and discussion}

The discovery of SM-like Higgs boson is a big progress for the particle physics and opens a new research era for the BSM.
In this short review, we focus on the topic of extra neutral and charged Higgs bosons at the LHC.

The charged Higgs is an ambiguous signature for the BSM. Utilizing the top polarization information,  $\tan{\beta}$ in the type-II 2HDM can be measured to a good precision, especially for the intermediate $\tan\beta$.  All of our analysis are based on tree-level estimation for the signal and background. The NLO QCD correction of $gb\rightarrow H^-t$  have been estimated long time ago~\cite{Zhu:2001nt,Baglio:2011ap}. Though the K-factor is roughly 1.8, the correction will not change our result significantly because the degree of top polarization is insensitive to higher-order correction. Our analysis is at parton-level, and the parton shower and hadronization can change the jets information. However for hard jets with large $p_T$ as required to suppress backgrounds, such effects should be insignificant.

The pair production of SM-like Higgs bosons can be the good probe for extra neutral Higgs boson, though the cross section in the SM  is small. The signal can be enhanced due to the resonance production
 of the extra neutral Higgs boson. Our analysis was carried out in effective Lagrangian. In fact such signal can exist in many BSM models, for example 2HDM for specific parameter set and the coloron model. Our numerical results show that the most promising modes are $HH \rightarrow b\bar b\gamma\gamma$ and $b\bar b\tau^-\tau^+$, and the less promising modes are$HH \rightarrow  b\bar bb\bar b$ and $b\bar b WW^*$. The detailed results can be found in
 table \ref{tab:discovery}.

In the BSM models, the extended Higgs sectors are usually required. Discovering such new particles and measuring their properties will be the interesting topic for
the LHC and future high energy colliders.

\section*{Acknowledgment}

This work was supported in part by the Natural Science Foundation
 of China (Nos. 11135003 and 11375014).

\bibliographystyle{JHEP}
\bibliography{referencelist}
\end{document}